\documentclass{aa}
\usepackage{graphicx}
\usepackage{txfonts}
\usepackage{natbib}
\bibpunct{(}{)}{;}{a}{}{,} 

\newcommand{\Msun}{\mbox{$M_{\odot}$}}

\begin{document}
\title{The pre-cataclysmic variable, LTT\,560}

\author{
C. Tappert\inst{1},
B. T. G\"ansicke\inst{2},
L. Schmidtobreick\inst{3},
A. Aungwerojwit\inst{2,4},
R. E. Mennickent\inst{5},
D. Koester\inst{6}
}

\authorrunning{C. Tappert et al.}

\offprints{C. Tappert}

\institute{
Departamento de Astronom\'{\i}a y Astrof\'{\i}sica, Pontificia Universidad
Cat\'olica, Vicu\~na Mackenna 4860, 782-0436 Macul, Chile\\
\email{ctappert@astro.puc.cl}
\and
Department of Physics, University of Warwick, Coventry CV4 7AL, UK\\
\email{Boris.Gaensicke@warwick.ac.uk, a.aungwerojwit@warwick.ac.uk}
\and
European Southern Observatory, Casilla 19001, Santiago 19, Chile\\
\email{lschmidt@eso.org}
\and
Department of Physics, Faculty of Science, Naresuan University, 
Phitsanulok, 65000, Thailand
\and
Departamento de F\'{\i}sica, Universidad de Concepci\'on, Casilla 160-C,
Concepci\'on, Chile\\
\email{rmennick@astro-udec.cl}
\and
Institut f\"ur Theoretische Physik und Astrophysik, University of Kiel,
24098 Kiel, Germany\\
\email{koester@astrophysik.uni-kiel.de}
}

\date{Received xxx; accepted xxx}

\abstract
{}
{
System parameters of the object LTT\,560 are determined in order to
clarify its nature and evolutionary status.
}
{
We apply time-series photometry to reveal orbital modulations of the
light curve, time-series spectroscopy to measure radial velocities of
features from both the primary and the secondary star, and flux-calibrated
spectroscopy to derive temperatures of both components.
}
{
We find that LTT\,560 is composed of a low temperature ($T \sim
7500$ K) DA white dwarf as the primary and an M$5.5\pm1$
main-sequence star as the secondary component.
The current orbital period is $P_{\rm orb} = 3.54(07)$ h. 
We derive a mass ratio $M_{\rm sec}/M_{\rm wd} = 0.36(03)$ and estimate 
the distance to $d = 25-40$ pc. 
Long-term variation of the orbital light curve and an additional
H$\alpha$ emission component on the white dwarf indicate 
activity in the system, probably in the form of flaring and/or
accretion events.  }
{}

\keywords{binaries: close -- Stars: late-type -- white dwarfs --
          Stars: individual: LTT 560 --
          cataclysmic variables}

\maketitle

\section{Introduction}

Cataclysmic Variables (CVs) are close interacting binaries with a 
white-dwarf (WD) primary accreting matter via Roche-lobe overflow from a
late-type companion that is on, or at least close to, the main sequence.
These systems are thought to form from initially detached binary stars
that go through a common-envelope (CE) phase 
\citep[ and references herein]{paczynski76-1}.
In the course of its nuclear evolution, the more massive star expands,
eventually fills its Roche lobe, and transfers matter onto the 
late-type dwarf at very high rates of up to 
$\dot{M}_1 \sim 0.01...0.1 M_{\sun} {\rm yr}^{-1}$ 
\citep[e.g.,][]{meyer+meyer-hofmeister79-1,warner95-1}.
The CE configuration results, since the secondary star cannot adjust
on such short timescales\footnote{Note, however, that neither high 
mass-transfer-rates lead necessarily to a CE, nor a CE phase to a CV 
\citep[e.g.,][ respectively]{beeretal07-1,dekool92-1}}.
The CE provides an enhanced braking mechanism, rapidly reducing the binary 
separation. After the expulsion of the CE, 
angular-momentum loss due to magnetic braking further reduces 
the orbital separation between the remaining WD and its red companion
(on much longer timescales). Eventually the Roche lobe of the red dwarf
shrinks into contact with the stellar surface, 
thus starting mass transfer and the proper CV phase. Following the
convention of \citet{schreiber+gaensicke03-1}, a post-CE binary is
called a pre-CV if the time between the expelling of the CE and the
start of the CV phase is less than the Hubble time.
For a comprehensive overview on CVs and CV evolution see \cite{warner95-1}.

Currently only $\sim40$ pre-CVs with established orbital
periods are known \citep{morales-ruedaetal05-1}, which is in strong
contrast to the over 600 CVs with measured orbital periods that are
listed in version 7.7 of the \cite{ritter+kolb03-1}
catalogue. Furthermore, this small sample is severely biased by
observational selection effects towards young systems containing hot
white dwarfs and low-mass companions
\citep{schreiber+gaensicke03-1}.

LTT\,560 is listed as a high proper motion system in the
\cite{luyten57-1} catalogue. The object was marked as a possible WD by
\cite{eggen68-1}, and appeared as a nova-like variable in
\cite{vogt89-1} and in \cite{downesetal97-1}, the latter using the
designation Scl2. Spectroscopy by \cite{hoard+wachter98-1} revealed
narrow H$\alpha$ emission and a red continuum that was fitted well by
an M5 dwarf. They also found a variable doubling of the
H$\alpha$ emission that led them to suspect a binary nature for LTT\,560.

Here we present photometric light curves and time-resolved and
flux-calibrated spectroscopy that reveal LTT\,560 as being an old and
nearby short-period pre-CV containing a very cool white dwarf.

\section{Observations and data reduction}

\begin{table*}
\caption[]{Log of observations. The configuration column for the 
spectroscopic data states the grism/grating, the wavelength range,
and the spectral FWHM resolution. Date and HJD refer to the start of night.}
\label{obs_tab}
\begin{tabular}{lllllll}
\hline\noalign{\smallskip}
date & HJD & telescope & configuration & $n_{\rm data}$ & $t_{\rm exp}$ [s]
& $\Delta t$ [h] \\
\hline\noalign{\smallskip}
2002-09-22 & 2\,452\,540 & 0.9 m & CCD+$R$ & 130 & 120 & 5.67 \\
2002-10-17 & 2\,452\,565 & 0.9 m & CCD+$R$ & 90 & 120 & 3.84 \\
2002-10-18 & 2\,452\,566 & 0.9 m & CCD+$R$ & 90 & 120 & 3.87 \\
2003-07-19 & 2\,452\,840 & NTT & g\#3, 3350--9000 {\AA}, 8.3 {\AA} 
& 3 & 400 & 0.25 \\
2004-11-29 & 2\,453\,339 & 4.0 m & KPGL3, 3500--7300 {\AA}, 4.2 {\AA}
& 27 & 600 & 5.57 \\
\hline\noalign{\smallskip}
\end{tabular}
\end{table*}

The photometric data were taken in two runs on 2002-09-22 and on 
2002-10-17+18 at the CTIO/SMARTS 0.9 m telescope using a 2048 $\times$ 2046
CCD and a Kron-Cousins $R$ filter. A scale of 0.396\arcsec/pixel yields a 
field-of-view of 13.5\arcmin. 
{The data were reduced using the quadred package implemented in 
IRAF\footnote{NOAO PC-IRAF Revision 2.12.2-EXPORT},
}
which takes into account that the CCD is read out via
4 amplifiers. Apart from that, the reduction process is standard, including
bias and overscan subtraction, as well as flatfielding. Aperture photometry
for all detected stars on the CCD was computed with IRAF's daophot package, and
additional daomatch and daomaster \citep{stetson92-1} routines were used for
cross-identification of stellar sources on different CCD frames. The adopted
aperture radius was chosen as the one that produced minimum noise in the
differential light curve of two comparison stars with brightness similar to
the target. An average light curve was computed out of several non-variable
comparison stars, and then used to produce differential light curves for all
stars in the field \citep[e.g., ][]{howell92-1}. 

Optical spectroscopy was performed on 2003-07-19 using EMMI on the NTT, at
ESO La Silla. Grism \#3 and a 1.0\arcsec slit yielded a wavelength range
of 3350--9000 {\AA} and a spectral resolution of 8.3 {\AA}. IRAF tasks were
used for basic reduction (bias + flats), optimal extraction \citep{horne86-1},
wavelength-, and flux calibration.

Time-resolved spectroscopic data were obtained on 2004-11-29 with the
R-C spectrograph mounted on the 4.0 m Blanco telescope at CTIO. The grating
KPGL3 was used with a 1.0\arcsec slit, which yielded a wavelength range of
3500--7300 {\AA} and a spectral resolution of 4.2 {\AA}. Wavelength calibration
data were taken with a HeAr lamp every 4 to 5 object spectra. 
The instrumental wavelength flexure between two calibration exposures was
later corrected for by measuring the position of the night sky emission line
at 6863 {\AA} and subtracting its variation with respect to the mean.
Weather 
conditions during this run were far from photometric, with a layer
of thin clouds during most of the time, so no flux calibration could be
performed. Data reduction was carried out as outlined above.

LTT\,560 was observed in the near-ultraviolet (NUV) and
far-ultraviolet (FUV) by GALEX as part of its All Sky Survey on
October 21, 2003. The 120 s exposures did not yield a detection in
the FUV, but provided an NUV flux of $0.236\pm0.01$\,mJy.

A summary of our observations is presented in Table \ref{obs_tab}.

\section{Results}

\subsection{The flux-calibrated spectrum\label{s-sedfit}}

\begin{figure}
\includegraphics[width=6.5cm,angle=270]{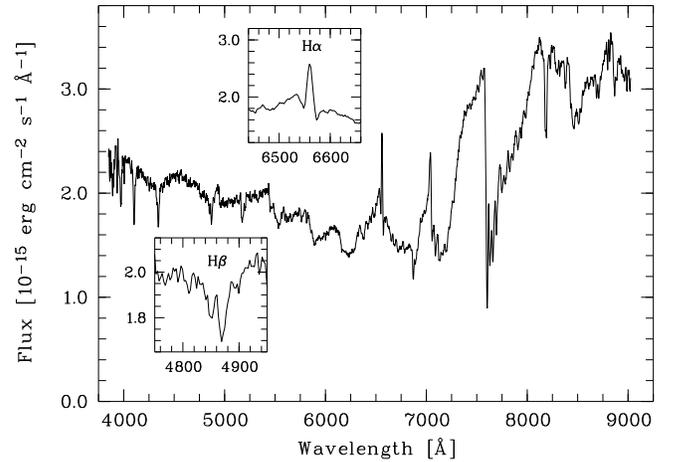}
\caption[]{The average flux-calibrated spectrum. The two insets show close-ups
of the regions around H$\beta$ and H$\alpha$.}
\label{avsp_fig}
\end{figure}

\begin{figure}
\includegraphics[angle=-90,width=\columnwidth]{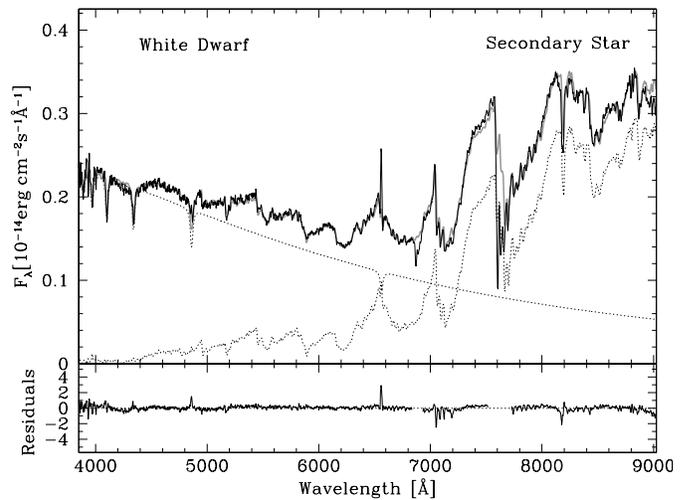}
\caption[]{Two-component fit (thick grey line) to the flux-calibrated
spectrum of LTT\,560 (solid black line). The
individual components, a white dwarf with a temperature of
$T_\mathrm{wd}=7500$\,K and a surface gravity of $\log g=8.0$, and 
our M5.5 template are shown as dashed lines.
The bottom panel shows the residuals of the fit, which
highlight the Balmer emission from the M-dwarf.}
\label{ltt560fit_fig}
\end{figure}

\begin{figure}
\includegraphics[angle=-90,width=\columnwidth]{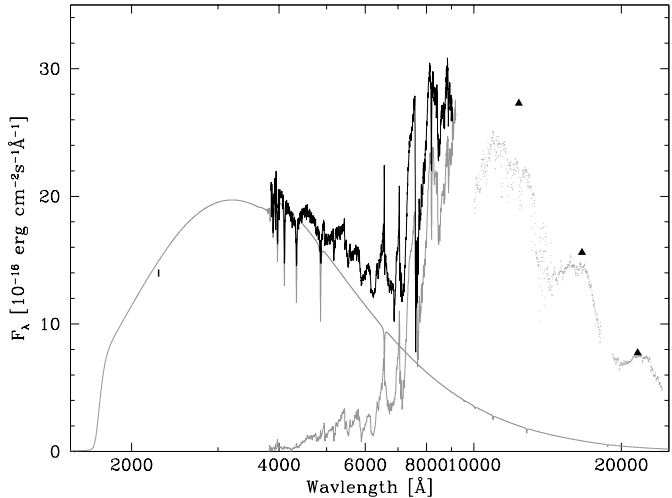}
\caption[]{Spectral energy distribution of LTT\,560. The GALEX 
near-ultraviolet flux, the CTIO average spectrum,
and the 2MASS $J$, $H$, and $K_S$ fluxes are shown in black. 
The $T_\mathrm{wd}=7500$\,K, $\log g=8.0$ white dwarf and the M5.5
M-dwarf from Fig.\,\ref{ltt560fit_fig} are plotted as grey lines. 
The grey dots represent an M5.5
template from S. Leggett's M-dwarf library, scaled according to our
fit to the optical data.}
\label{sed_fig}
\end{figure}

The three individual spectra taken on 2003-07-19 were averaged to
yield the spectrum displayed in Fig.\ \ref{avsp_fig}. It shows a blue
continuum with Balmer absorption lines, and a red continuum with
molecular absorption bands. The relatively narrow width of
the Balmer absorption lines suggests a cold
($T_\mathrm{eff}<10\,000$\,K) white dwarf, and the strength of the TiO
bands an intermediate M-type secondary star. On closer inspection, the
presence of narrow Balmer emission is clearly seen in H$\alpha$ and
H$\beta$. Convolving the spectrum with \cite{bessell90-1} filter
curves yields spectrophotometric magnitudes $V = 15.7$ and $R = 15.2$
mag. This is roughly 0.2\,mag brighter than previously reported
\citep[ and references therein]{hoard+wachter98-1}, but such small
difference can probably be attributed to the uncertainties of the flux
calibration, and we re-scaled our spectrum of LTT\,560 to $V=15.7$ for
the subsequent analysis.

We fitted a composite model consisting of a synthetic white dwarf
spectrum and an observed M-dwarf template to the flux-calibrated
spectrum of LTT\,560. The synthetic white dwarf spectra were
calculated with the code described by \citet{koesteretal05-1} and
earlier references given there. We used a mixing length
ML/2$\alpha=0.6$, a pure hydrogen composition, a surface gravity $\log
g=8.0$, corresponding to a white dwarf mass of
$M_\mathrm{wd}\simeq0.55$\,\Msun, and covered a temperature range
$T_\mathrm{wd}=6000$\,K to 15\,000K.  
Our library of high
signal-to-noise ratio M-dwarf templates covers spectral types
M0--M9. These templates were produced by averaging about a dozen
high-quality SDSS spectra of each spectral type, where the typing of
the SDSS spectra was done based on set of carefully calibrated M-dwarf
spectra from \citet{beuermann+weichhold99-1}. The advantage of our new
set of templates over that of \citet{beuermann+weichhold99-1} is a
homogenous wavelength coverage and spectral resolution for all
spectral types. A comparison of our M-dwarf templates to those
recently published by \citet{bochanskietal07-1} showed very good
agreement for all spectral types. We scaled the M-dwarf templates to
the surface fluxes corresponding to their spectral type using the
relation of \citet{beuermann06-1}.  The best fit to the spectrum of
LTT\,560 is achieved for a $T_\mathrm{wd}=7500$\,K white dwarf and
either an M5 or M6 secondary star. Figure\,\ref{ltt560fit_fig}
shows the observed spectrum of LTT\,560 along with the white dwarf
model, and an M5.5 star obtained from linear interpolation between the
M5 and M6 templates, with both components scaled according to the best
fit.  Earlier and later spectral types for the secondary star result
in much poorer fits, and we conclude that the spectral type of the
main-sequence companion in LTT\,560 is $5.5\pm1$. This result is 
confirmed by recent $K$-band spectroscopy \citep{tappertetal07-2}.

Our fit to the optical spectrum of LTT\,560 agrees well with the
2MASS magnitudes of LTT\,560 ($J=12.65$, $H=12.13$, and
$K_\mathrm{s}=11.86$) when extending the composite model into the
infrared with Leggett's library of M-dwarf
spectra\footnote{http://ftp.jach.hawaii.edu/ukirt/skl/dM.spectra/}
(Fig.\,\ref{sed_fig}). The best-fit white dwarf model spectrum overpredicts 
the GALEX near-UV flux only by a few per cent. Given the uncertainty in the
spectrophotometric calibration of our optical data, and the fact that
the 2MASS data have been taken at unknown binary phases, an attempt to
fit the overall spectral energy distribution would not result in more
accurate stellar parameters, and has therefore not been attempted.

Assuming a \citet{hamada+salpeter61-1} carbon-core mass-radius
relation for the white dwarf, the flux scaling factor between the
synthetic spectrum and the data of LTT\,560 implies a distance of
$d=33$\,pc. This estimate depends somewhat on the white dwarf
mass. Repeating the fit for $\log g=7.5$ ($M_\mathrm{wd}=0.33$\,\Msun)
and $\log g=8.5$ ($M_\mathrm{wd}=0.90$\,\Msun) results in temperatures
lower and larger by 500 K, respectively. The implied distance is 39\,pc
for $\log g=7.5$ and 26\,pc for $\log g=8.5$,  and we conclude
$25\,\mathrm{pc}\la d\la 40$\,pc. This makes the object a good target
for a ground-based parallax measurement.

For $d=33$\,pc, the flux scaling factor of our
M-dwarf spectral template implies a radius of the secondary of
$R_\mathrm{sec} \sim 8.0\pm1.3\times10^9$\,cm, which is consistent with an 
$\sim$M5V secondary star \citep{delfosseetal00-1, segransanetal03-1}. 
For a mass ratio of
$M_\mathrm{sec}/M_\mathrm{wd}=0.32$ (Sect.\,\ref{s-rv}) and an orbital
period of 3.54 h (Sect.\,\ref{s-lc}), the Roche lobe radius of the
secondary star is $\sim2.1\times10^{10}$\,cm, which implies that the
secondary star is $\sim50$\% Roche-lobe filling.  
Given the amplitude of the observed ellipsoidal modulation
(Fig.\,\ref{f-lightcurves}), the inferred radius is clearly too
small. This will be discussed further in Sect.\,\ref{disc_sec}.

\subsection{The light curve\label{s-lc}}

\begin{figure}
\includegraphics[width=\columnwidth]{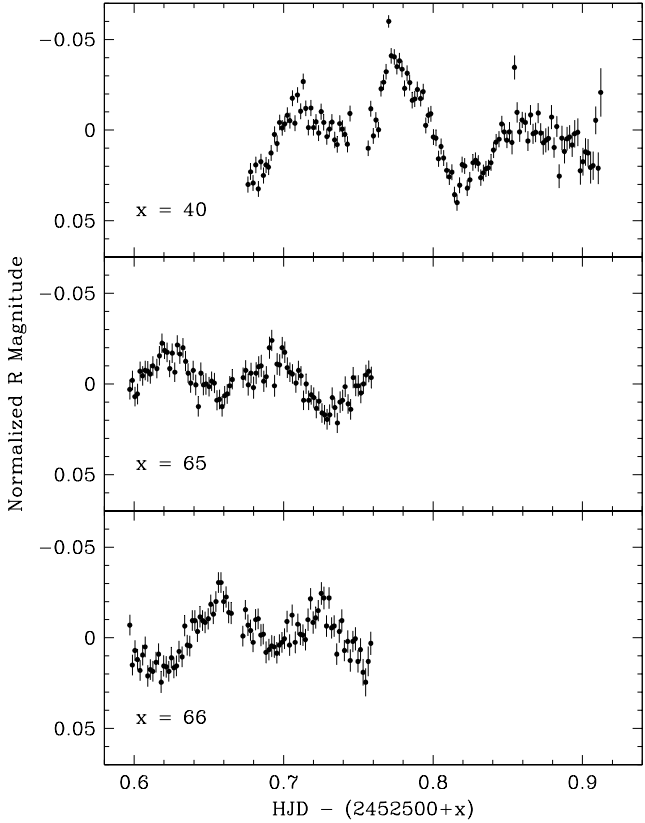}
\caption[]{\label{f-lightcurves} Individual light curves from 2002-09-22, 
2002-10-17, and 2002-10-18 (top to bottom). The data have been normalised
by subtracting the average magnitude of the complete data set.}
\label{indlc_fig}
\end{figure}

\begin{figure}
\includegraphics[angle=270,width=\columnwidth]{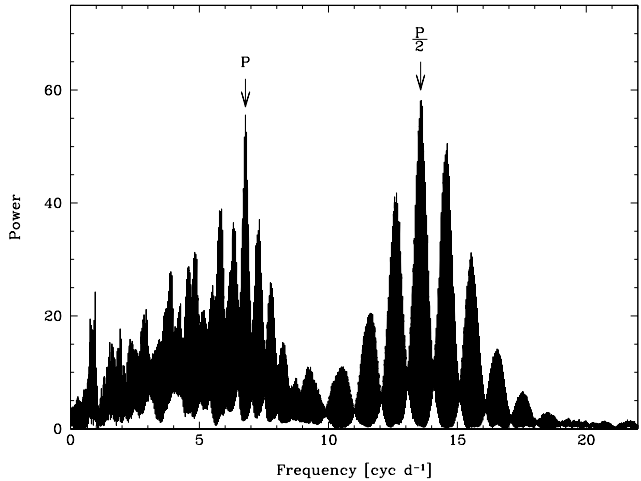}
\caption[]{AOV periodogram of the photometric data. The best period and
half its value are indicated.}
\label{pgram_fig}
\end{figure}

\begin{table}
\caption[]{Timings of suitable photometric minima.}
\label{min_tbl}
\begin{tabular}{lll}
\hline\noalign{\smallskip}
cycle & HJD (2452500+) & error [d]\\
\hline\noalign{\smallskip}
0.0 & 65.6575 & 0.0097 \\
0.5 & 65.7316 & 0.0055 \\
7.0 & 66.6937 & 0.0064 \\
\hline\noalign{\smallskip}
\end{tabular}
\end{table}

\begin{figure}
\includegraphics[angle=270,width=\columnwidth]{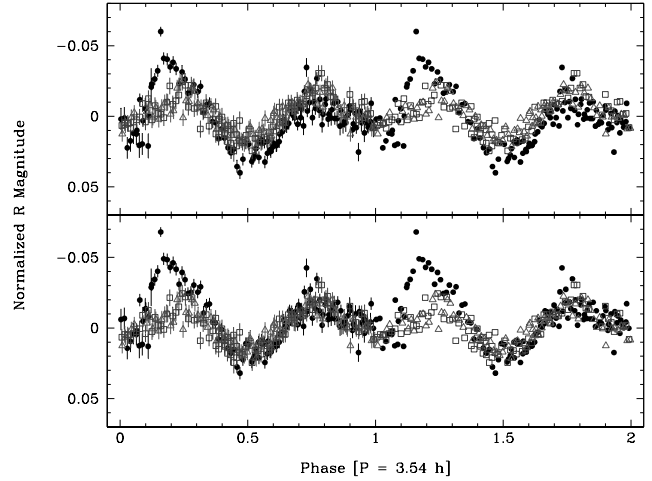}
\caption[]{Light curves folded on the ephemeris in Eq.(\ref{ephem_eq}).
Filled circles, triangles, and squares, mark data from 2002-09-22, 2002-10-17, 
and 2002-10-18, respectively. The data are repeated, without error bars, 
for a second cycle. For the lower plot, an offset of $-0.008$ mag has been 
applied to the data from 2002-09-22.}
\label{philc_fig}
\end{figure}

The resulting light curves of the two observing runs from September and in
October 2002 are given in Fig.\,\ref{indlc_fig}. All three data sets show
as a common feature the presence of two minima of different depths,
indicating ellipsoidal variation. In the first data set, from September 2002,
the two maxima have different values as well.

To explore the periodicity of the light curve, the analysis-of-variance
\citep[AOV; ][]{schwarzenberg-czerny89-1} algorithm as implemented in
MIDAS was used. The two highest peaks in the periodogram 
(Fig.\,\ref{pgram_fig}) correspond to frequencies $f_1 = 6.786$ cyc d$^{-1}$ 
and $f_2 = 13.609$ cyc d$^{-1}$, and thus $f_2 \approx 2 f_1$. We therefore
adopt the lower frequency as corresponding to the orbital period, and describe 
the photometric variation as ellipsoidal: the two minima represent the inferior
and superior conjunction of the late-type secondary star, the two maxima
are the result of maximum visibility of the Roche-deformed sides of that star 
at quadrature.
From $f_1$ we derive a corresponding period $P = 3.54(07)$ h, with the
uncertainty being estimated from the width of the signal in the periodogram.
Fig.\,\ref{philc_fig} presents the phase-folded light curve.
In principle, there are two effects that can account for the difference in the
minima: at superior conjunction either illumination by the white dwarf 
causes an elevated minimum, or gravitational darkening of the deformed
secondary star results in a deeper minimum. Considering the very low 
temperature of the white dwarf, the latter scenario appears as the most
probable one. 

Since the orbital phase is now defined we can go on to derive the
photometric ephemeris. In the light curve there are three well defined
minima that can be fitted with second-order polynomials in order to derive
the timings of these minima. These include both the shallow and the deep
minimum from 2002-10-17, and the shallow minimum from 2002-10-18. A linear
fit to the such derived timings (Table \ref{min_tbl}) yields the zero point
and the orbital period. The latter results in 3.55(03) h, and thus, within
the errors, is identical to the one derived from the periodogram. However, 
since it is based on only three data points, the formal error of 0.03 h is
probably underestimated. We therefore adopt the period and uncertainties 
estimated from the periodogram, to obtain the ephemeris 
\begin{equation}
T_0 ({\rm HJD}) = 2\,452\,565.6575(51) + 0.1475(29)~E,
\label{ephem_eq}
\end{equation}
with $E$ giving the cycle number with respect to the superior conjunction
of the white dwarf.

We note that, while the two light curves from October are identical
within the photometric errors, the September data set not only presents an
enhanced second maximum, but on the whole appears to be displaced to
$\sim$0.01 mag fainter magnitudes. 
Comparison of the average
magnitudes of common field stars in the September and the October run
yields a standard deviation of 0.008 mag, so that this displacement
can be explained as being within the photometric precision limit with
respect to these two runs. If a corresponding offset of
$-0.008$ mag is applied to the September data, the general shape of
that light curve is identical to the October data, with the
notable exception of the second maximum (lower plot of
Fig.\,\ref{philc_fig}). 
The latter feature is discussed in more detail in
section \ref{disc_sec}.

We used the program \texttt{Nightfall} written by
Wichmann\footnote{http://www.lsw.uni-heidelberg.de/$\sim$rwichman/Nightfall.html}
to test the parameters derived from the spectroscopic model
(Sect.\,\ref{s-sedfit}) by fitting synthetic light curves to the
October 2002 data. A first light curve was calculated adopting the parameters
from Table\,\ref{tab_rvfit}, in particular $P_\mathrm{orb}=0.1475$\,d,
$i=60\degr$, $T_\mathrm{wd}=7500$\,K, $T_\mathrm{sec}\simeq3000$\,K,
$M_\mathrm{wd}\simeq0.4-0.6$\,\Msun, $M_\mathrm{sec}=0.17$\,\Msun,
$R_\mathrm{wd}\simeq0.012-0.016\,R_\odot$, and
$R_\mathrm{sec}=0.115\,R_\odot$. However, the ellipsoidal modulation of the
resulting light curve is of much lower amplitude ($\sim0.0015$) than the
observed one. We then fitted the observed light curve by fixing all
parameters above but leaving $R_\mathrm{sec}$ as a free parameter. We find 
that, to reproduce an ellipsoidal modulation corresponding to
the observed light curve, the required $R_\mathrm{sec}$ exceeds the 
spectroscopic result by a factor of two. The discrepancy
between the spectrophotometric radius (Sect.\,\ref{s-sedfit}) and that
implied by the ellipsoidal modulation is significant and will be discussed in
Sect.\,\ref{disc_sec}.

\subsection{\label{s-rv}Radial velocities}

\begin{figure}
\includegraphics[angle=-90,width=\columnwidth]{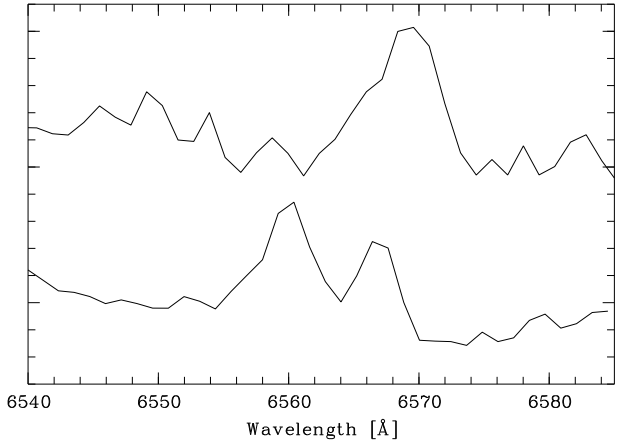}
\caption[]{The H$\alpha$ line profile at the time of maximum separation of
the two emission components (bottom), and half an orbital cycle later (top).}
\label{Haprof_fig}
\end{figure}

\begin{figure}
\includegraphics[width=\columnwidth]{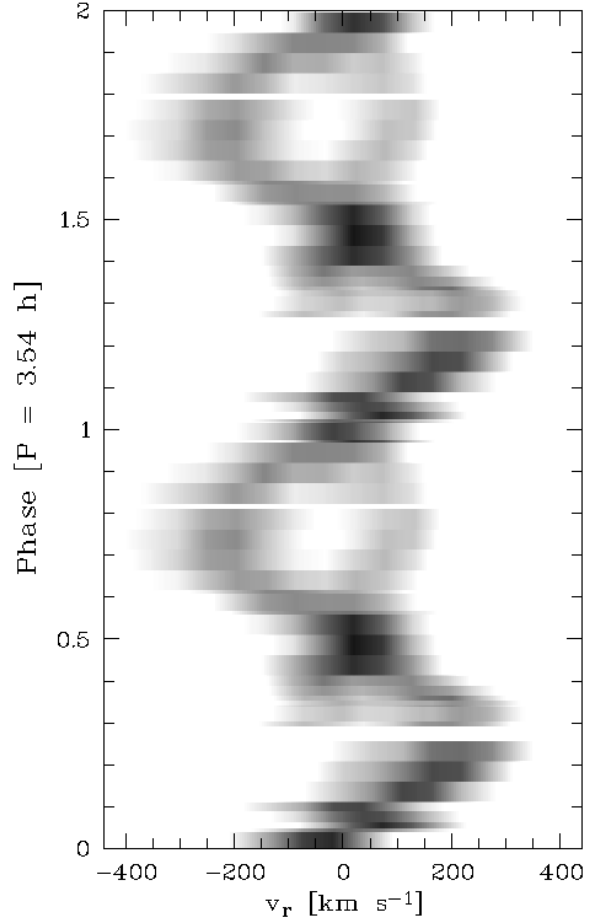}
\caption[]{Trailed spectrogram of the 2004-11-29 data, folded on 
$P = 3.54$ h. Radial velocities refer to the rest wavelength of H$\alpha$.
The data are repeated for a second cycle.}
\label{trsp_fig}
\end{figure}

The H$\alpha$ emission line is clearly double-peaked during
substantial parts of the orbital cycle (Fig.\,\ref{Haprof_fig}). The trailed 
spectrogram (Fig.\,\ref{trsp_fig}) reveals that the two components follow 
sinusoidal radial velocity variations with a phase difference of $\simeq0.5$.  
To measure the radial velocity variation, we fitted the spectra that present 
two clearly separated
H$\alpha$ components with two Gaussians, finding that these components are
unresolved at the spectral resolution of the CTIO spectra. 

\begin{figure}
\includegraphics[angle=-90,width=\columnwidth]{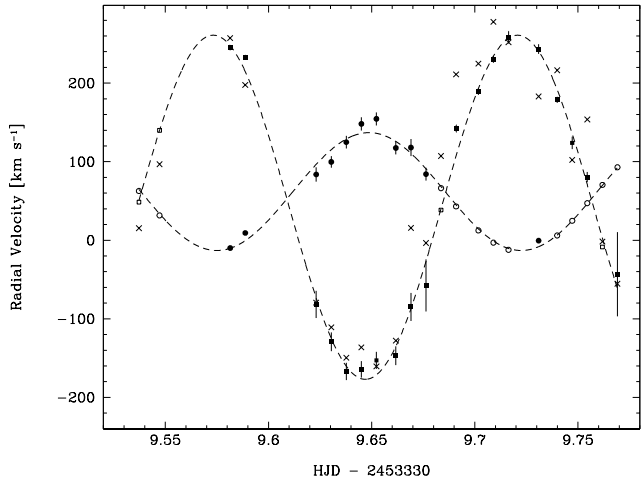}
\caption[]{Radial velocities for both H$\alpha$ components.
For components marked with filled symbols, the velocities were
obtained by simultaneous Gaussian fits to both $H\alpha$ emission line
components. In some cases the weaker component was ill-defined, and
the two-Gaussian fit was carried out with the position of the weaker
component fixed to the wavelength predicted by a sine fit to the velocities
obtained from the spectra where a two-Gaussian fit was possible. Those
points with fixed velocities are marked by open symbols. Four spectra
were too poor to obtain any significant fit, and are marked by open
symbols for both components. Additional radial velocities extracted from
the TiO absorption edge at 7040 {\AA} are marked by $\times$. These data were
not included in the fit, but are overplotted for comparison.}
\label{harvboth_fig}
\end{figure}

Next, we attempted to fit all individual spectra with two
Gaussians of $\mathrm{FWHM}=4.2$\,\AA, i.e. with their widths fixed to
the spectral resolution of our data, leaving only their position and
amplitude free. For a number of spectra the fit failed to disentangle
the two line profiles, especially during those orbital phases when
both components cross, and for the spectra from the last third of our
run, which suffered from degrading conditions, and where the weaker
H$\alpha$ component was ill-defined. In the cases where one component
was still clearly discernable, we fitted the combined profile with two
Gaussians as described above, but fixed the position of the weaker
component to the velocity predicted by a sine-fit to the radial
velocities from the spectra where both components could unambiguously
be disentangled. For four spectra, no meaningful result could be
extracted from the line profiles. The radial velocity curves derived
in this way are shown in Fig.\,\ref{harvboth_fig}.

The two radial velocity variations appear sinusoidal in shape, and are
offset by half a cycle. We fitted both sets of radial velocities with
sine curves of the form $K\sin[2\pi(t-T_0)/P]+\gamma$ with $K$ the
radial velocity amplitude, $t$ the time of the observation, $T_0$ an
offset to account for the arbitrary phase during the observations,
$P=0.1475$ d the orbital period determined from the photometry, and a
constant velocity offset $\gamma$. The fit parameters are reported in
Table\,\ref{tab_rvfit}.

In Fig.\,\ref{harvboth_fig}, we include radial velocities that 
were derived by fitting a single Gaussian to the absorption edge of the
TiO molecular band near $\lambda7040$ {\AA}. 
A sine fit yields the following
parameters for this component: $\gamma = 89 \pm 5$ km/s, $K = 208 \pm 7$ km/s,
$T_0 ({\rm HJD})= 2\,453\,339.534 \pm 0.001$, which both in amplitude and
phase agrees well with the high-amplitude H$\alpha$ component, which is also
the stronger of the two components.

We therefore ascribe the two lines moving in anti-phase to emission
from the companion star and the white dwarf. Coronal activity in
late-type stars is quite common, especially if forced to rapid
rotation by being tidally locked in a short-period binary.  However,
emission from the white dwarf is rather unusual, and may indicate a
corona related to low-level accretion from the M-star wind
\citep[see][ for a similar phenomenon in the pre-CV EC 13471-1258]{odonoghueetal03-1}. 
Taking such an origin of the two emission lines at face value, their radial
velocity variations trace the motion of both stars, and the ratio of
their amplitudes equals the mass ratio of the two stars,
$K_\mathrm{wd}/K_\mathrm{sec}=M_\mathrm{sec}/M_\mathrm{wd}=0.36\pm0.03$.
We note no substantial change in the equivalent width of the H$\alpha$
emission line from the companion star, which implies that the motion
of the centre of light is equivalent to the motion of the centre of
mass. Comparing the velocity offsets obtained from the two fits, it
becomes apparent that $\gamma_\mathrm{wd}$ exceeds
$\gamma_\mathrm{sec}$ by 25\,km/s. Assuming that $\gamma_\mathrm{sec}$
reflects the radial component of systemic velocity, we interpret
$\gamma_\mathrm{wd}-\gamma_\mathrm{sec}$ as the gravitational redshift
of the H$\alpha$ emission originating on the white dwarf. General
relativity predicts a redshift of
$v_\mathrm{gr}=0.635(M/M_\odot)/(R/R_\odot)$, and assuming again a
\citet{hamada+salpeter61-1} mass-radius relationship,
$v_\mathrm{gr}=25\pm9$\,km/s corresponds to
$M_\mathrm{wd}=0.52\pm0.12\,M_\odot$. A caveat to this mass
measurement is that if the H$\alpha$ emission arises from a point
somewhat above the white dwarf photosphere
(\citealt{odonoghueetal03-1} find that the second component of the H$\alpha$ 
emission in EC 13471-1258 has its origin somewhere between the white dwarf 
and L$_1$), its mass would be underestimated.  
The mass of the secondary star implied by the mass
ratio above is $M_\mathrm{sec}=0.19\pm0.05~\Msun$.
If the companion star
follows the spectral type-mass-radius relation of typical low-mass
stars \citep{segransanetal03-1, delfosseetal00-1} the masses of both
stars are likely to be on the lower side of the error ranges. Given
the fact that the mass ratio and both radial velocity amplitudes are
known, we can estimate the binary inclination as $\simeq60^{\circ}$.

\begin{table*}
\caption{\label{tab_rvfit} System parameters of LTT\,560.}
\begin{tabular}{cccl}
\hline\noalign{\smallskip}
                 & white dwarf     & main sequence star & method\\
\hline\noalign{\smallskip}
Orbital period [h] & \multicolumn{2}{c}{$3.54\pm0.07$\,h} & light curve\\
Inclination        & \multicolumn{2}{c}{$\simeq60^\circ$} & radial velocities\\
Distance [pc]      & \multicolumn{2}{c}{$25-40$\,pc} & spectrophotometry\\
$K$ [km/s]       & $80\pm6$        & $220\pm9$  & radial velocities\\
$T_0$ [d]        & $0.566\pm0.002$ & $0.052\pm0.001$ & radial velocities\\
$T_0$ [HJD 2\,453\,339 +] & $0.463\pm0.002$ & $0.538\pm0.001$ 
& radial velocities\\
$\gamma$ [km/s]  & $95.0\pm5.2$    & $70.0\pm6.7$ & radial velocities \\
$T_\mathrm{eff}$ / spect. type & $7500\pm500$\,K & M5$\pm0.5$ 
& spectrophotometry \\
Mass [$M_\odot$] & $0.52\pm0.12$   & $0.19\pm0.05$ & radial velocities\\
Radius [cm]      & $(9\pm1)\times10^8$ & $(8\pm1.3)\times10^9$ 
& spectrophotometry \\
\hline\noalign{\smallskip}
\end{tabular}
\end{table*}

\section{\label{disc_sec}Discussion}

In comparison with other known pre-CVs, LTT\,560 represents a special
case in that it shows spectroscopic evidence for low-level accretion,
and apparently irregular long-term variability in its light
curve. 
Accretion from the wind of the companion star has been
observed in other pre-CVs in the form of metal enrichment of the white
dwarf photosphere or additional emission components 
\citep[e.g.,][]{sionetal98-2, obrienetal01-1, odonoghueetal03-1, 
maxtedetal07-1}. \citet{schmidtetal05-1,schmidtetal07-1} also ascribe the
mass transfer mechansim in low accretion rate polars 
\citep[LARPS,][]{schwopeetal02-1} to wind from the secondary star.
Concerning the light curve, flaring activity from the
secondary star appears as the most likely explanation for the enhanced
photometric maximum in the September data, since the H$\alpha$
component from the secondary star indicates the presence of
chromospheric activity, and the duration of the ``flare'' feature with
$\sim$0.15 orbits $\approx$30 min is within the usual time range of
such flares \citep[e.g.,][]{garciaalvarez00-1}. Still, since the
orbital phase in question is only covered once in the September data,
it remains unclear whether this feature is not a periodic event,
representing an additional light source in the system, perhaps due to
enhanced accretion at that time.

A weak X-ray source at a count rate of $0.008\pm0.002$\,cts/s was 
serendipitously detected in a pointed ROSAT PSPC observation that coincides 
within the positional error with LTT\,560 \citep{whiteetal00-1}. Adopting a 
distance of 33\,pc and the countrate-to-flux conversion factor of 
\citet{flemingetal93-1}, we find an X-ray luminosity of 
$\simeq6\times10^{27}$\,erg/s, which is consistent with coronal activity on 
the M-dwarf. 

A point worth further discussion is the disagreement between the
spectrophotometric radius of the secondary star
(Sect.\,\ref{s-sedfit}) and that implied by the observed amplitude of
the ellipsoidal modulation (Sect.\,\ref{s-lc}).  
A well-known
problem in the understanding of low mass stars is that the radii of
stars in eclipsing binaries with masses of $\simeq0.4-0.8\,M_\odot$
are larger by up to 15\% than predicted by stellar models
(\citealt{ribas06-1} for a review), and stellar activity is generally
seen as a possible explanation for this discrepancy
\citep[e.g.,][]{lopez-morales07-1}. However, radii of low-mass
stars from interferometric measurements also appear to be too large
\citep{bergeretal06-1}. In the context of pre-CVs, the companion star
is tidally forced into rapid rotation, which is believed to result in
increased stellar activity \citep{pallavicinietal81-1,
pizzolatoetal03-1}. Whether or not stellar activity is the cause of
anomalies in the radii, masses, and spectral types of the companion
stars in pre-CVs \citep[e.g.,][]{obrienetal01-1, maxtedetal04-1} or CVs
\citep[e.g.,][]{nayloretal05-1} is not yet clear, as no fully detailed
treatement of the structure and the observational properties of
strongly active low-mass stars is available. In a pioneering paper,
\citet{spruit+weiss86-1} showed that for stars above the fully
convective boundary, both the radius and the temperature of the star
will be underestimated if a large fraction of the star is covered by star
spots. However, the effects of magnetic activity are thought to extend
to lower masses as well \citep{mullan+macdonald01-1}, which is
confirmed by observations \citep{westetal04-1}. The calculations of
Mullan \& MacDonald
demonstrated that for low mass M-dwarfs, magnetic activity may result
in effective temperatures that are too low for their radii. While a
full discussion has to await progress in the theoretical models of
active stars, it appears possible that activity caused by rapid
rotation induces anomalies in the relationships between mass, radius,
and spectral type, and that this is the cause of the apparent
mis-match of the spectrophotometric radius of the companion star in
LTT560, and its radius implied by the ellipsoidal modulation.

The uncertainty in the stellar properties of the companion
star propagates into a relatively wide range of possible periods when
LTT\,560 will evolve into a semi-detached configuration (i.e. start
mass transfer as a cataclysmic variable). For example, using equations
(8) and (11) in \citet{schreiber+gaensicke03-1}, and assuming that
gravitational radiation is the main mechanism for angular momentum loss
($M_\mathrm{sec} < 0.3~M_\odot$) yields $P_\mathrm{cd} = 0.165~\mathrm{d}$, 
$P_\mathrm{sd} = 0.035~\mathrm{d} \sim 50~\mathrm{min}$ (which is not a very
probable value for a CV), $t_\mathrm{sd} = 2.9~\mathrm{Gyr}$, for
$R_\mathrm{sec} = 0.115~R_\odot$, and $P_\mathrm{sd} = 0.093~\mathrm{d}
\sim 113~\mathrm{min}$, $t_\mathrm{sd} = 2.0~\mathrm{Gyr}$, for 
$R_\mathrm{sec} = 0.22~R_\odot$, with $P_\mathrm{cd}$ being the orbital period 
at the end of the CE phase, $P_\mathrm{sd}$ the orbital period at which 
the system will start mass transfer (i.e., become a CV), and 
$t_\mathrm{sd}$ the time the system will need to shrink from the current 
orbital period to $P_\mathrm{sd}$. 
Additionally, the spectral type derived in Sect.\,\ref{s-sedfit} suggests 
$P_\mathrm{sd}<2$ h, while the mass ratio determined from the radial 
velocities and the gravitational redshift indicates $P_\mathrm{sd}>3$h. 
An accurate parallax determination of LTT\,560 would therefore be extremely 
useful for a better constraint on the stellar parameters and the future 
evolution of LTT\,560.

The white dwarf cooling age of LTT\,560, assuming
$M_\mathrm{wd}\simeq0.5\,M_\odot$, is $\simeq10^9$\,yr
\citep{wood95-1}. The relatively low mass suggests that it may contain
a helium core, which is a possible outcome of the common envelope
evolution (\citealt{willems+kolb04-1} and references therein). Better
data are needed to improve the measurement of the white
dwarf mass.

LTT\,560 and RR\,Cae \citep{bragagliaetal95-1, bruch+diaz98-1,
bruch99-1} represent the oldest among the $\sim40$ known post-common
envelope binaries \citep{schreiber+gaensicke03-1,
morales-ruedaetal05-1, shimanskyetal06-1}.  Both systems are nearby
($\simeq$33\,pc and $\simeq$12\,pc, respectively) which suggests that
they represent a fairly numerous population, as predicted to exist by
\citet{schreiber+gaensicke03-1}.  
A principle difficulty in
finding old pre-CVs is their inconspicuousness compared to CVs. Both
stellar components in old pre-CVs are intrinsically faint, and hence
only nearby systems will be found in shallow surveys. Pre-CVs do not
undergo outburst, so will not be picked up by amatuer astronomers of
sky patrols. They have at best some X-ray emission due
to coronal activity of the main sequence companion, indistinguishable
from single M-dwarfs. Old pre-CVs are not
particularly blue, so escaped discovery in the classic ``blue''
extragalactic surveys such as the Palomar Green Survey.  The Sloan
Digital Sky Survey (SDSS), being deeper and sampling a much larger
colour space than any previous survey, is now rapidly changing the
scene, identifying already $\ga1000$ white dwarf main sequence
binaries \citep{silvestrietal06-1, eisensteinetal06-1,
southworthetal07-1}, with more to come from SEGUE
\citep{schreiberetal07-1}. However, follow-up work identifying
post-CE/pre-CV binaries among those objects has just started
\citep[e.g.][]{rebassa-mansergasetal07-1}, and will be limited (by the
design of SDSS) predominantly to relatively faint systems. While SDSS
will substantially improve our knowledge on pre-CVs in terms of better
statistics, identifying nearby bright pre-CVs such as LTT\,560 white
dwarf main sequence binaries remains important as those systems are
best-suited for detailed parameter studies.

\begin{acknowledgements}
This research has greatly benefited from discussions with Tom Marsh, Matthias 
Schreiber, Raymundo Baptista, and Stella Kafka. We also thank the anonymous
referee for helpful suggestions.

CT and RM acknowledge financial support by FONDECYT grant 1051078.  BTG was
supported by a PPARC Advanced Fellowship.  AA thanks the Royal Thai
Government for a studentship. 
This work has made intensive use of the SIMBAD database, operated at CDS, 
Strasbourg, France, and of NASA's Astrophysics Data System Bibliographic 
Services. IRAF is distributed by the National Optical Astronomy Observatories.
\end{acknowledgements}


\end{document}